# Dual regimes of ion migration in high repetition rate femtosecond laser inscribed waveguides

T Toney Fernandez[1*], B. Sotillo[2], J. del Hoyo[1], J. A Valles[3], R. Martinez Vazquez[4], P. Fernandez[2], J. Solis[1*]

[1]*Grupo de Procesado por Láser, Instituto de Optica, CSIC, Serrano 121, 28006-Madrid, Spain*
[2] *Departamento de Física de Materiales, Facultad de Físicas, Univ. Complutense, 28040-Madrid, Spain*
[3]*Departamento de Física Aplicada – I3A, Facultad de Ciencias, Universidad de Zaragoza, Zaragoza, Spain*
[4]*Instituto di Fotonica e Nanotecnologie del CNR – Dipartimento di Fisica del Politecnico di Milano, Milano, Italy*

*Abstract -* **Ion migration in high repetition rate femtosecond laser inscribed waveguides is currently being reported in different optical glasses. For the first time we discuss and experimentally demonstrate the presence of two regimes of ion migration found in laser written waveguides. Regime-I, corresponds to the initial waveguide formation mainly via light element migration (in our case atomic weight < 31u), whereas regime-II majorly corresponds to the movement of heavy elements. This behavior brings attention to a problem which has never been analyzed before and that affects laser written active waveguides in which active ions migrate changing their local spectroscopic properties. The migration of active ions may in fact detune the pre-designed optimal values of active photonic devices. This paper experimentally demonstrate this problem and provides solutions to avert it.**

## I. INTRODUCTION

ION MIGRATION in high repetition rate femtosecond laser written waveguides[1] is being currently reported in many glasses[2-5]. This paper focuses on how the migration of ions affects the active performance of the fabricated waveguides. A waveguide optical amplifier is modelled for its optimum performance in terms of dopant concentration, total length of the waveguide, absorption coefficient of the bulk glass, etc [6-8]. A fabrication technique should thus, in principle, not affect the active spectroscopic properties of the base material (practically, a subtle modification could be expected). Conventional techniques, like Ag-Na ion exchange, RIE ion exchange, are considered as reliable as they normally don't modify the active spectroscopic properties [7, 9].

Since the introduction of femtosecond laser writing technique we have seen a tremendous growth in the number of devices that have been fabricated both in 2 and 3 dimensions. Indeed waveguide amplifiers and lasers, fabricated by ultrafast laser inscription, provide in many cases a performance comparable to that exhibited by devices fabricated by ion exchange or plasma-enhanced chemical vapor deposition [10]. Active device fabrication by femtosecond laser inscription has been particularly interesting for materials that are unmanageable to be processed by other fabrication techniques, eg: Tellurites, Tellurides, Chalcogenides etc [11, 12].

Focusing on the results of C-band amplifiers fabricated by both ultrafast laser inscription and Ag-Na ion exchange, surprisingly femtosecond laser inscription technique has not always matched the benchmarks already set by Ag-Na ion exchange process. For example, an optical amplifier fabricated by ion-exchange demonstrated a record 5.4 dB/cm performance (net gain per unit length) in a 3.1 cm long waveguide doped with 2.3 wt% of $Er^{3+}$ and 3.6 wt% of $Yb^{3+}$. However, for a 2 wt% of $Er^{3+}$ and 4 wt% of $Yb^{3+}$, the best achieved net gain result upon fs-laser writing was 2.5 dB/cm from a 3.7 cm long waveguide (3.5dB/cm from a 2 cm long waveguide) [13]. Several explanations were put forward. The most widely accepted one is the strong asymmetry of the refractive index profile produced by laser writing, disturbing the mode field overlap between pump and signal [7].

Recently we have observed and reported strong ion migration effects in high repetition rate (500kHz and 1MHz) ultrafast laser written waveguides. In these works the tendency of multivalent ions to move in opposite directions with respect to monovalent ions has been observed. In particular, the first ($La^{3+}$, $Al^{3+}$, $Yb^{3+}$, $Er^{3+}$, $Ce^{3+}$, $Zn^{2+}$, $Si^{2+,4+}$, $Te^{4+, 6+}$) move towards the densified zone while the monovalent ions ($K^+$, $Na^+$, $P^{+, 3+, 5+}$) move towards the rarefied zone of the waveguide [3, 4]. The term 'valence' here is solely used to classify the ions based on their common valences. The observed migration of multivalent ions in a Lanthanum Phosphate glass lead to local compositional changes of some elements from 3 % to 33 % relative to their initial concentration [2].

In this paper, for the first time, we experimentally report on the effects and potential problems derived from active ion migration in photonic devices fabricated by high repetition rate laser inscription, something that has been overlooked over the past years. We provide solutions to avert this situation using both simulations and experimental data. The paper also puts forward a new experimental observation indicating two regimes of material densification caused by the migration of light and heavy elements, respectively.

## II. EXPERIMENTAL

The waveguides were laser written using a high repetition rate femtosecond laser (Tangerine, Amplitude Systems) operating at a wavelength of 1030 nm. Waveguides in this paper were written with a repetition rate of 500 kHz using a 0.68 NA aspheric lens. The composition of the glass used in this work is $64P_2O_5.10La_2O_3.10Al_2O_3.8K_2O.2SiO_2$ (mol%) doped with $Er_2O_3$ (2 wt.%) and $Yb_2O_3$ (4 wt. %) The energy dispersive X-ray micro-analysis measurements of the waveguides were carried out in a Leica S440 SEM equipped with a Bruker AXS Quantax micro-analysis system. The refractive index mapping was carried out at 670 nm using a

Manuscript received December X, XXX. Corresponding authors: *toney.teddyfernandez@io.cfmac.csic.es and *j.solis@io.cfmac.csic.es
Digital Object Identifier inserted by IEEE

Rinck Elektronik refractive index profilometer.

### III. RESULTS AND DISCUSSION

The role of ion migration in high repetition rate femtosecond laser inscribed waveguides was recently reported [3, 4]. These reports described its role on the local material densification leading to a high refractive index region and thus enabling light guiding. In the experiment, a set of writing pulse energies were used to produce densification within the phosphate glass. The DIC microscope image along with the secondary electron SEM image of waveguides written within an energy window from 520 nJ to 700 nJ are shown in figure 1. In both types of images, the white contrast is indicative for densification. The EDX compositional profiles for two representative waveguides, clearly illustrating two different ion migration regimes, are shown in figure 2.

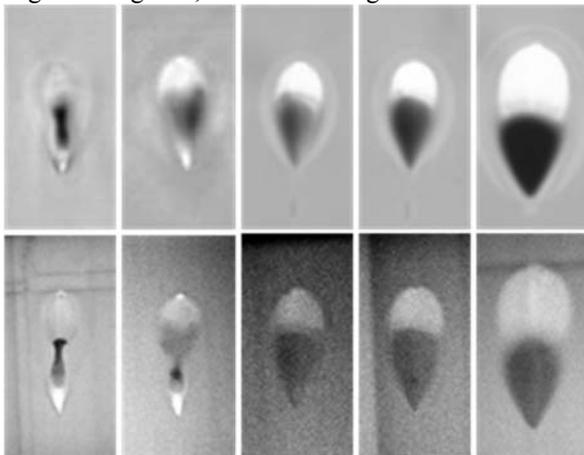

Fig. 1. DIC images (1st row) and SE images (2nd row) of waveguides written at 520, 550, 580, 610, 700 nJ (*40 μm x 20 μm image sizes*)

Aluminum and silicon play a major role in the material densification when waveguides are written at low energies (low energy ion migration regime (regime-I, <600nJ)). It is worth noting that the potassium concentration underwent a massive 81 % increase in the rarefied zone for a waveguide written with energy of 520 nJ whereas for 700 nJ inscribed waveguide the potassium migration was only 30 %. Since a higher defect density below or near the damage threshold is expected, the driving force for the migration of monovalent ions is most likely the presence of defects. In regime-I, though heavy elements also show migration, it lacks a definite behavior as found in the high energy regime-II. At higher energies (regime-II), heavy elements are activated indicating a definite behavior and role in material densification, whereas light elements go almost inert or lack a definite trend. This clearly indicates the existence of an activation energy threshold for individual elements in a multicomponent glass for ultrafast laser-matter interaction. The explanation for this behavior in a multicomponent glass involves strong complexity. In any case this thresholding behavior has been reported elsewhere providing proper explanations with respect to the laser induced plasma distributions [14, 15]. In these works it is also shown that for energies in the ~610-730 nJ interval, the pulse energy is essentially invested in the La-K cross-migration process rather than increasing the volume of the structure[3].

Ion migration has in fact helped to open the bottle necking associated to the refractive index change (Δn) achievable in high repetition rate femtosecond inscribed waveguides. The obtained Δn was between $5 \times 10^{-3}$ to $5 \times 10^{-2}$ [15], as shown in figure 3a where a near step-index profile can be appreciated due to a near saturated contribution from La ions (c.f. Fig 2b, lanthanum profile). A non-saturated profile should infact produce a graded index profile due to the ion migration enabling the tuning of the refractive index profile pattern also, but this completely depends on the glass composition and its constitutents. These Δn values, along with the use of a simple slit shaping technique, enabled to perform an excellent V-number tuning and thus matching the waveguide parameters to those of a Corning SMF-28 fiber (figure 3b). Though V-number is a parameter often used to represent the essential properties of a step index fiber, we have used it here for the sake of simplicity, to explain the dependence between the number of modes, Δn and waveguide dimensions. The waveguides were designed to operate for $Yb^{3+}$ pumping (978 nm) and C-band signal enhancement (1520-1565 nm) using $Er^{3+}$ doping.

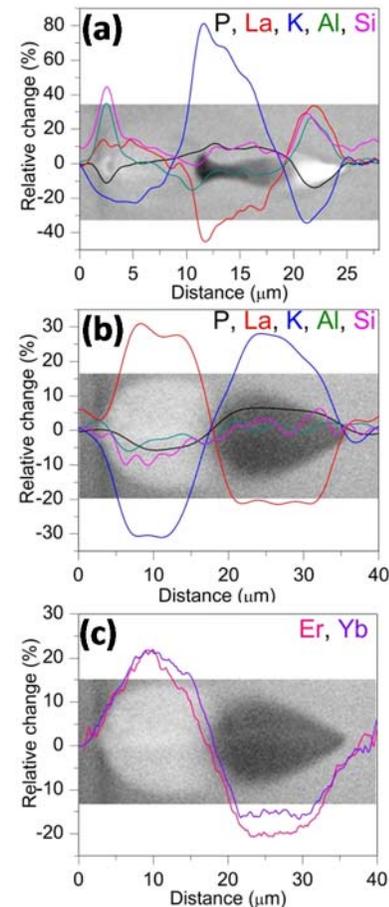

Fig. 2. Ion migration observed in two waveguides written at **(a)** 520 nJ **(b, c)** 700 nJ.

First as a positive implication, the refractive index was varied by controlling the ion migration via pulse energy and the dimension of the waveguide was varied by using slits [13]. Hence the tuning of V-number of the waveguides eventually resulted in finally matching the parameters to a Corning SMF-28 fiber[16]. Figure 4 shows the overall picture of the propagated mode characteristics in waveguides written with different energy and slits. The single mode propagation was extended over a wider energy window offering a larger degree of freedom in waveguide writing. This total procedure is in fact a mimicking of Ag-Na ion exchange process [17] where the role of the metal negative masks (fabricated by sputtering deposition and selective etching) was replaced by laser beam slit shaping and the alkali ion concentration of the salt bath and ion exchange time was replaced by laser fluence (which in turn controls the ion migration).

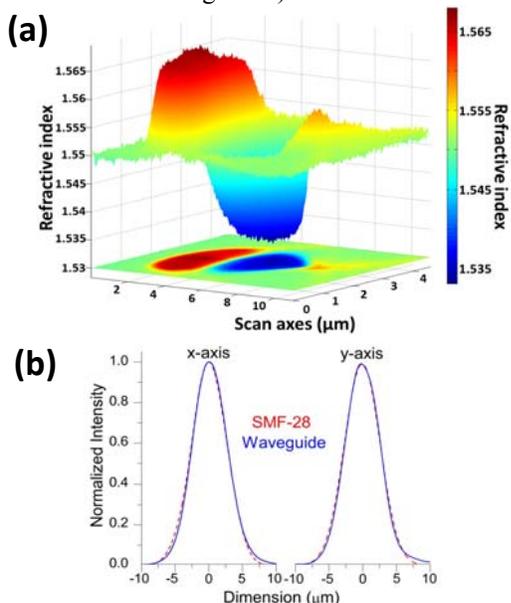

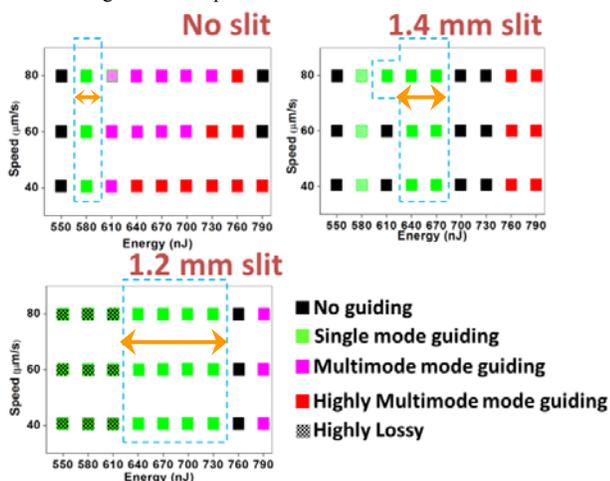

Fig. 3. (a) Refractive index profile of the waveguide written with laser energy of 700 nJ. (b) 1620 nm mode from the waveguide in comparison with that of an SMF-28 fiber.

Fig. 4. Effect of slit beam shaping and ion migration in tuning the V-number of the waveguides.

In the higher energy regime, the migration of active ions like erbium and ytterbium (Figure 2c), is quite noticeable upon the laser energy increase. This is, to the best of our knowledge, the first report on the migration of RE active ions ($Er^{3+}$, $Yb^{3+}$) upon high repetition rate laser waveguide writing and is, in principle, a detrimental side-effect for pre-designed active photonic devices. To analyze its impact in an optical amplifier design, we have selected a set of waveguides showing gradual increase in Erbium and Ytterbium content.

The experimental quantification of active ion migration for each waveguide is given in Table 1 along with the observed optical absorption at 1534 nm ($^4I_{15/2} \rightarrow ^4I_{9/2}$), signal enhancement and internal gain due to this effect. The first row corresponds to a simulated value calculated from the spectroscopic parameters of the unirradiated bulk sample. The optical amplification experiments were carried for bidirectional pumping (420 mW) in 2 cm-long waveguides.

In agreement with the local $Er^{3+}$ concentration increase ($Yb^{3+}$ concentration also increases by the same amount, cf. Fig 2c), there is a large increase of absorption, enhancement, and internal gain reaching a value of 9.4 dB for 2.1 cm, which in turn could be proposed in a range of applications like miniature loss-less splitters, amplifiers or rare-earth based sensors, which one might think is a favorable situation. It has to be considered though that the situation corresponds to a non-optimized sample length. Hence, we have carried out several simulations [18] to show the effect of active ion migration over the pre-designed values of an optical amplifier (Figure 5).

| TABLE I | | | | |
|---|---|---|---|---|
| Energy (nJ) | Relative increase of $Er^{3+}$(%) | Absorption (dB) for 2.1 cm | Enhancement (dB) | Internal gain (dB) |
| - | 0 | 9.2 | 17.2 | 8 |
| 610 | 8.6 | 9.9 | 15.4 | 5.5 |
| 700 | 21.8 | 11 | 18.8 | 7.8 |
| 730 | 23.25 | 11.3 | 19.5 | 8.2 |
| 760 | 23.85 | 11.4 | 19.7 | 8.3 |
| 790 | 24 | 12.4 | 21.8 | 9.4 |

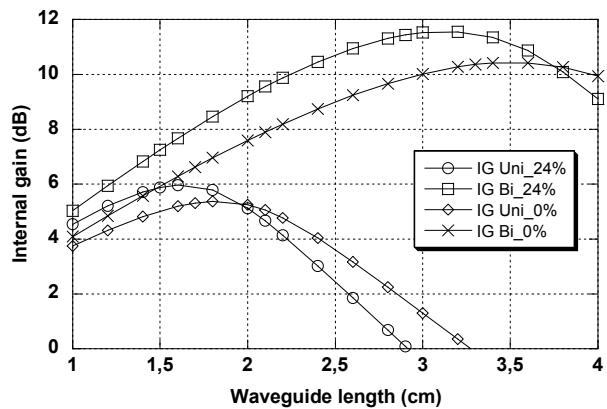

Fig. 5. Simulation of internal gain vs length showing the effect of ion migration. (IG - internal gain, Uni – Unidirectional pumpimg 210mW, Bi-Bidirectional pumping, 420 mW, 0% - No ion migration, 24% - local concentration increase of $Er^{3+}$)

The simulation (figure 5) clearly indicates that the pre-designed optimum length value for this lanthanum phosphate glass with a doping of 2 wt% $Er^{3+}$ and 4 wt% $Yb^{3+}$ was 3.6 cm with an internal gain value of 10.5 dB for bi directional pumping. The experimental optimal length value was shifted to 3 cm with a maximum value of 12 dB. Apparently positive in this case, in general, ion migration has to be carefully treated in the device optimization. For instance, a 3.5 wt% $Er^{3+}$ and 7 wt% $Yb^{3+}$ phosphate glass [19] would end-up absorbing 0.85 dB/mm and a variation in device length of just 5mm would increase the absorption by 4-5 dB, reducing the gain. Moreover, for low phonon energy glasses like tellurites, tellurides and chalcogenides, the effect will be much more extended, as these glasses have poor energy transfer rate coefficient [20]. In the case of doubly or triply codoped glasses for cascaded energy transfer [21], the effect poses a great challenge in designing a stable operation. As a consequence, RE active ion-migration effects in waveguides produced by high repetition fs-laser writing should be carefully considered and analyzed when producing the end products. This can be done by modeling the expected impact of the ion migration in a realistic manner, as shown in Fig.5.

A possible solution to avoid detuning with respect to the pre-designed values of the laser written waveguide optical amplifier is to work at regime-I, where only light elements and glass matrix changes are responsible for densification. This will be at the cost of a lower refractive index change but this is not a great concern for active waveguides. Also, as shown by fig.5, it is possible to include in the simulations of device design the expected enrichment of active ions in the guiding region.

## IV. CONCLUSION

We have experimentally verified that there exist two regimes of ion migration upon high repetition rate fs-laser irradiation of multicomponent glasses. In Regime-I, for waveguides written at low energies, light elements are responsible for the refractive index modification. Within regime-II heavier elements are responsible for waveguide formation. It is noticeable that ion migration under high repetition rate fs-laser processing has to be taken as an expected effect beyond the specific glass composition. An important consequence is that, for regime-II, the active rare-earth ions also migrate which is an undesired result. We analyzed this observation and found that even for a moderately doped phosphate glass, ion migration caused an alteration of the pre-designed values such as optimal length, absorption within the waveguide, and internal gain. We can thus infer that active ion migration has to be carefully considered in the design of actvie devices to be produced by high repetition rate fs-laser writing.

ACKNOWLEDGEMENTS

Work partially supported by the Spanish Ministry of Economy & Competitiveness (MINECO, TEC2011-22422, MAT2012-31959), J.H. & T.T.F. acknowledge funding from the JAE CSIC Program (co-funded by the European Social Fund). B Sotillo acknowledges her funding in the frame of CSD2009-00013 (MINECO).

REFERENCES

[1] S. Eaton, H. Zhang, P. Herman, F. Yoshino, L. Shah, J. Bovatsek, *et al.*, "Heat accumulation effects in femtosecond laser-written waveguides with variable repetition rate," *Optics Express,* vol. 13, pp. 4708-4716, Jun 2005.

[2] M. Shimizu, M. Sakakura, M. Ohnishi, M. Yamaji, Y. Shimotsuma, K. Hirao, *et al.*, "Three-dimensional temperature distribution and modification mechanism in glass during ultrafast laser irradiation at high repetition rates," *Optics Express,* vol. 20, pp. 934-940, Jan 2012.

[3] T. Toney Fernandez, P. Haro-González, B. Sotillo, M. Hernandez, D. Jaque, P. Fernandez, *et al.*, "Ion migration assisted inscription of high refractive index contrast waveguides by femtosecond laser pulses in phosphate glass," *Optics Letters,* vol. 38, pp. 5248-5251, Dec 2013.

[4] T. T. Fernandez, M. Hernandez, B. Sotillo, S. M. Eaton, G. Jose, R. Osellame, *et al.*, "Role of ion migrations in ultrafast laser written tellurite glass waveguides," *Optics Express,* vol. 22, pp. 15298-15304, Jun 2014.

[5] P. Mardilovich, L. Yang, H. Huang, D. M. Krol, and S. H. Risbud, "Mesoscopic photonic structures in glasses by femtosecond-laser fashioned confinement of semiconductor quantum dots," *Applied Physics Letters,* vol. 102, 151112, Apr 2013.

[6] J. A. Valles, "Method for Accurate Gain Calculation of a Highly $Yb^{3+}/Er^{3+}$ Codoped Waveguide Amplifier in Migration-Assisted Upconversion Regime," *Quantum Electronics, IEEE Journal of,* vol. 47, pp. 1151-1158, Jun 2011.

[7] G. D. Valle, "Photonic devices at 1.5 microns manufactured by ion exchange and femtosecond laser writing," PhD Thesis, Dipartimento di Fisica, Politecnico di Milano, Milano, 2007.

[8] J. A. Valles, A. Ferrer, J. A. Sanchez-Martin, A. R. de la Cruz, M. A. Rebolledo, and J. Solis, "New Characterization Technique for Femtosecond Laser Written Waveguides in Yb/Er-Codoped Glass," *Quantum Electronics, IEEE Journal of,* vol. 46, pp. 996-1002, Jun 2010.

[9] I. Bányász, I. Rajta, G. U. L. Nagy, Z. Zolnai, V. Havranek, S. Pelli, *et al.*, "Ion beam irradiated optical channel waveguides," Proc. SPIE, vol 8988, pp 898814-1 - 898814-9, Mar 2014.

[10] R. Osellame, G. Cerullo, and R. Ramponi, *Femtosecond laser micromachining: photonic and microfluidic devices in transparent materials* vol. 123. Berlin, Germany: Springer-Verlag, 2012.

[11] K. Vu and S. Madden, "Tellurium dioxide Erbium doped planar rib waveguide amplifiers with net gain and 2.8dB/cm internal gain," *Optics Express,* vol. 18, pp. 19192-19200, Aug 2010.

[12] K. T. Vu and S. J. Madden, "Reactive ion etching of tellurite and chalcogenide waveguides using hydrogen, methane, and argon," *Journal of Vacuum Science & Technology A: Vacuum, Surfaces, and Films,* vol. 29, pp. 011023-011023-6, Jan 2011.

[13] R. Osellame, N. Chiodo, G. Della Valle, G. Cerullo, R. Ramponi, P. Laporta, *et al.*, "Waveguide lasers in the C-band fabricated by laser inscription with a compact femtosecond oscillator," *Selected Topics in Quantum Electronics, IEEE Journal of,* vol. 12, pp. 277-285, May 2006.

[14] T. T. Fernandez, J. Siegel, J. Hoyo, B. Sotillo, P. Fernandez, and J. Solis, "Controlling plasma distributions as driving forces for ion migration during fs laser writing," *Journal of Physics-D:Applied Physics (In press),* March 2015.

[15] J. del Hoyo, R. M. Vazquez, B. Sotillo, T. T. Fernandez, J. Siegel, P. Fernández, *et al.*, "Control of waveguide properties by tuning femtosecond laser induced compositional changes," *Applied Physics Letters,* vol. 105, 131101, Sep 2014.

[16] W. Yang, C. Corbari, P. G. Kazansky, K. Sakaguchi, and I. C. Carvalho, "Low loss photonic components in high index bismuth borate glass by femtosecond laser direct writing," *Optics Express,* vol. 16, pp. 16215-16226, Sep 2008.

[17] R. V. Ramaswamy and R. Srivastava, "Ion-exchanged glass waveguides: a review," *Lightwave Technology, Journal of,* vol. 6, pp. 984-1000, Jun 1988.

[18] J. A. Vallés, A. Ferrer, J. M. Fernández-Navarro, V. Berdejo, A. Ruiz de la Cruz, I. Ortega-Feliu, *et al.*, "Performance of ultrafast laser written active waveguides by rigorous modeling of optical gain measurements," *Optical Materials Express,* vol. 1, pp. 564-575, Jun 2011.

[19] J. Hoyo, V. Berdejo, T. T. Fernandez, A. R. A Ferrer, J. A. Valles, M. A. Rebolledo, *et al.*, "Femtosecond laser written 16.5 mm long glass-waveguide amplifier and laser with 5.2 dB cm−1 internal gain at 1534 nm," *Laser Physics Letters,* vol. 10, p. 105802, Oct 2013.

[20] A. Jha, B. Richards, G. Jose, T. Teddy-Fernandez, P. Joshi, X. Jiang, *et al.*, "Rare-earth ion doped $TeO_2$ and $GeO_2$ glasses as laser materials," *Progress in Materials Science,* vol. 57, pp. 1426-1491, Nov 2012.

[21] S. Xu, H. Ma, D. Fang, Z. Zhang, and Z. Jiang, "$Tm^{3+}/Er^{3+}/Yb^{3+}$-codoped oxyhalide tellurite glasses as materials for three-dimensional display," *Materials Letters,* vol. 59, pp. 3066-3068, Oct 2005.